\documentclass[10pt]{article}
\usepackage{wws2023} % Uses Times Roman font (either newtx or times package)
\usepackage{url}
\usepackage{latexsym}
\usepackage{amsmath, amsthm, amsfonts}
\usepackage{algorithm, algorithmic}  
\usepackage{graphicx}
\usepackage{lipsum} 
\usepackage{fancyhdr}
\usepackage[headheight=1in,margin=1in]{geometry}
\usepackage{hyperref}
\usepackage[utf8]{inputenc}
\hypersetup{colorlinks=true,citecolor=[rgb]{0,0.4,0}}

\title{Synia: Displaying data from Wikibases}

\author{
  Finn Årup Nielsen \\
  Technical University of Denmark
  \\}

\begin{document}
\maketitle
\thispagestyle{fancy}

\begin{abstract}
I present an agile method and a tool to display data from Wikidata and other
Wikibase instances via SPARQL queries.
The work-in-progress combines ideas from the Scholia Web application
and the Listeria tool. 
\end{abstract}


\section*{Introduction}

Scholia is a Web application running from the Wikimedia Foundation
Toolforge server at \url{http://scholia.toolforge.org}.
It displays data from Wikidata via SPARQL queries to the Wikidata
Query Service (WDQS), particularly showing metadata about
scientific publications \cite{Q41799194}, chemical information
\cite{Q53844513}, and software \cite{Q113992928}.
The Web application is implemented with the Python Flask framework and
SPARQL templates are defined with Jinja2 templates that are read
during the application startup and interpolated based on the Scholia
user browsing.
Two other tools use a similar Flask/SPARQL template approach to display
Wikidata data:
Ordia is specialized for the lexicographic part of Wikidata
\cite{Q63862226} and CVRminer\footnote{\url{https://cvrminer.toolforge.org/}.} on Danish companies.
Common limitations for these tools are currently
\begin{enumerate}
\item The tools are bound to the Wikidata WDQS endpoint
\item The language is fixed to English
\item Development of new panels and aspects requires the involvement of
  software developers.
\end{enumerate}

For Magnus Manske's Listeria tool, wiki editors define MediaWiki templates with SPARQL queries on wikipages.
The Listeria bot then edits on behalf of the user and generate tables
on the wikipage according to the SPARQL query.\footnote{\url{https://listeria.toolforge.org/}.}

The approach I will describe here was first explored in a specific
instance of a Wikibase for data related to environmental impact
assessment reports \cite{Nielsen2023Environmental}.
In this abstract, I describe the extension of the approach, so it can
be used more widely with only slight changes in configurations in and across
different Wiki\-bases, --- including Wikidata.

\section*{Methods}

I call the tool \emph{Synia} with the canonical homepage set up at \url{https://synia.toolforge.org/}.
The implementation is a serverless single-page application (SPA)
consisting of a simple HTML page and some JavaScript.
Instead of storing the SPARQL templates along with the Web
application,
the templates are stored on wikipages.
The URL pattern of Scholia is borrowed and changed to use URI
fragments to control which wikipage should be read and what values
should be interpolated in the template.
Table~\ref{tab:mapping} shows some of the mapping between the URI
fragment and the wikipage.
A pseudo-namespace, Wikidata:Synia, is used as the default for
grouping the templates.
If the template is not defined on the wiki Synia creates a link, so a user/editor can
create the template.
Faceted search is supported, e.g.,
``\#venue/Q15817015/topic/Q2013'' shows information about the topic
\emph{Wikidata} occurring in the journal \emph{Semantic Web}.
Aspects with multiple items, e.g., hand\-ling
``\#authors/Q20980928,Q20895241,Q20895785'' is not yet supported.

\begin{figure}[ht]
  \begin{center}
    \centerline{\includegraphics[width=\columnwidth]{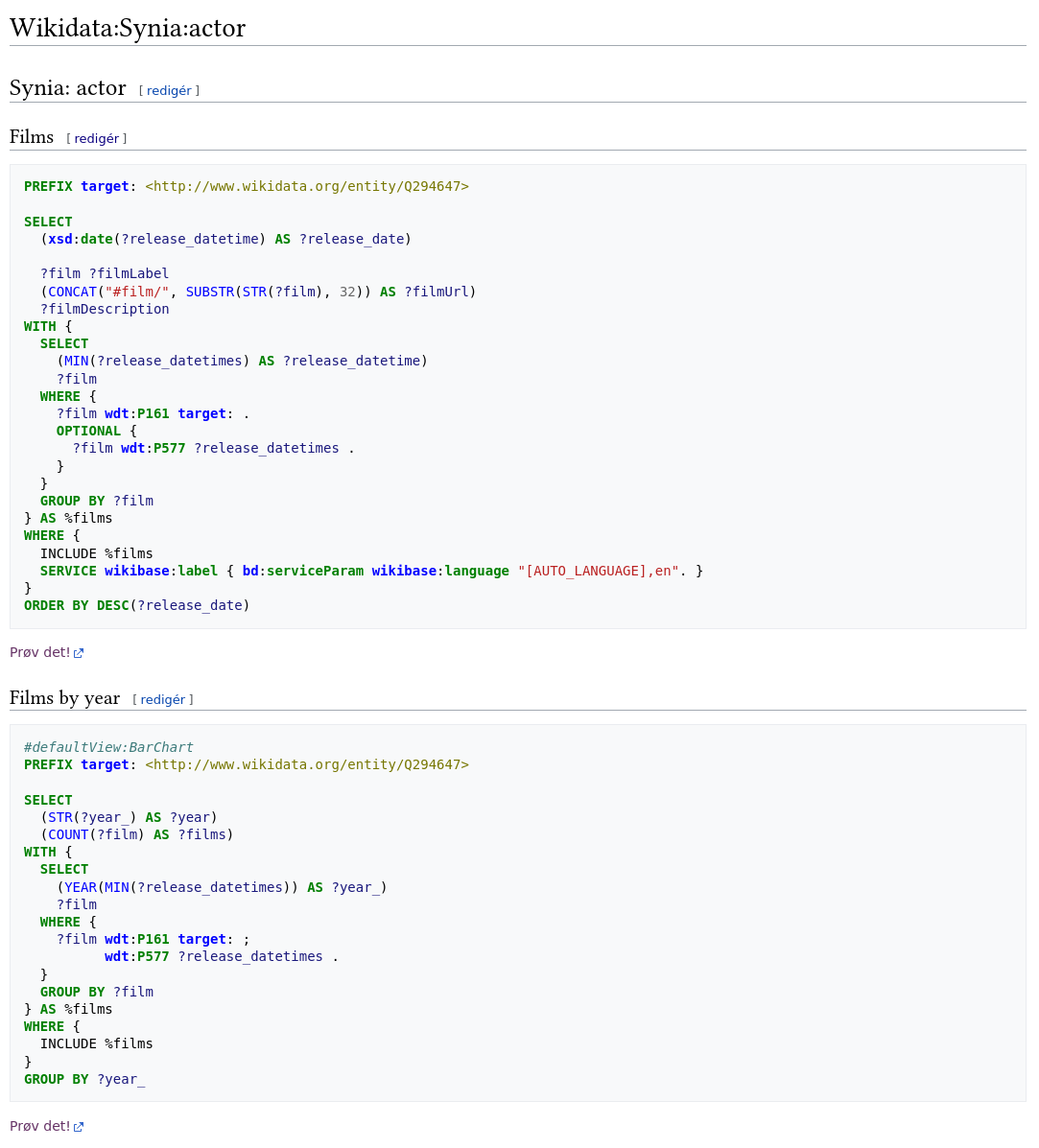}}
    \caption{Actor template at \url{https://www.wikidata.org/wiki/Wikidata:Synia:actor}.}
    \label{fig:template}
  \end{center}
\end{figure}

When wikipages are used for templates there are at least two important
issues to consider:
The template should be humanly readable as a wikipage and the
information read should be untrusted as wikis are usually openly
editable.
Currently, a limited set of components are handled, see
Table~\ref{tab:components}.
The parsing of the components is based on a series of regular
expressions. 
Synia will recognize MediaWiki headings and render them with h1, h2,
and h3 HTML tags.
SPARQL templates for Synia are stored on the wikipage in the
\emph{Template:SPARQL} MediaWiki template.
Synia extracts the SPARQL code, interpolates the Q- and L- identifier(s), and
sends the interpolated SPARQL to the SPARQL endpoint.
The response is rendered as a table in the SPA using the DataTables
JavaScript library or it may be rendered as a graph in an iframe with
the graphing capabilities of the query service.
For the ordinary wiki user, the template wikipage appears as ordinary
wikipages with SPARQL as code examples, see Figure~\ref{fig:template}.
The wikipage may have multiple headings and SPARQL templates.

Other endpoints than the configured default can be queried.
Currently Synia abuses an \emph{endpoint} parameter for the
\emph{Template:SPARQL} Media\-Wiki template on Wikidata to specify the
other endpoint.
An example using the approach is currently displayed at
\url{https://www.wikidata.org/wiki/Wikidata:Synia:compound} where a
panel for a SPARQL query goes to the endpoint of the
\url{https://wikifcd.wikibase.cloud} wiki \cite{Q113573682}.
This wiki has a Wikidata mapping property, so the Q-identifier can be
matched across Wikibases to a Wikidata identifier.

Bootstrap, jQuery, and DataTables libraries are used.
To avoid leaking browsing behavior the static files are hosted along
with the SPA.
Configuration, e.g., about the location of templates and the default
endpoint is maintained in a separate JavaScript file.

% \section*{Results}

\begin{figure}[ht]
  \begin{center}
    \centerline{\includegraphics[width=\columnwidth]{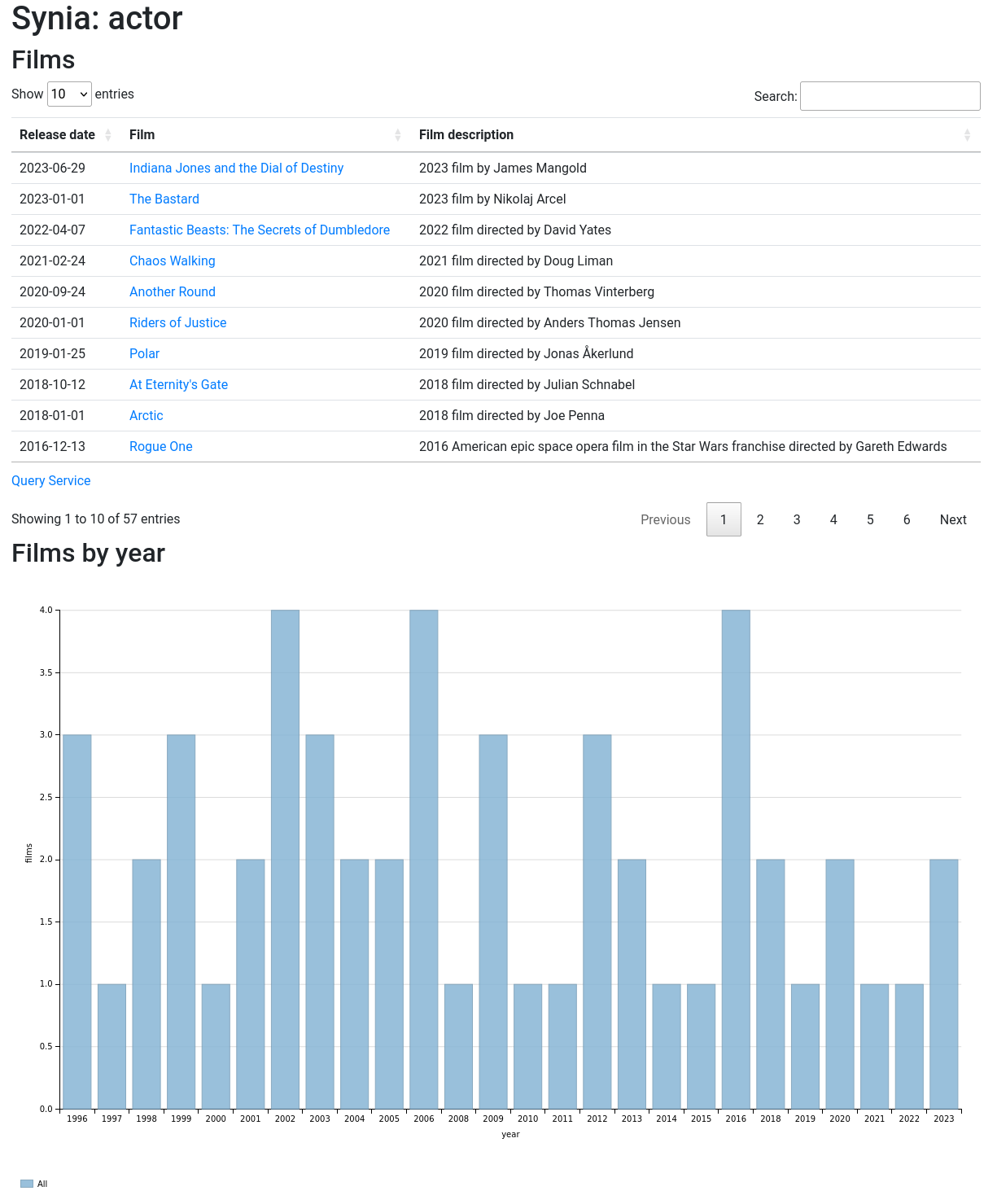}}
    \caption{Rendered page for Wikidata entity Q294647 in the actor
      aspect of Synia at \url{https://synia.toolforge.org/\#actor/Q294647}.}
    \label{fig:syniaactor}
  \end{center}
\end{figure}

A few aspects have so far been defined for Synia each with a few panels, e.g.,
author, work, venue, film, actor, compound, and lexeme.
Figure~\ref{fig:syniaactor} shows a screenshot of the actor aspect for
the Wikidata entity
\href{https://www.wikidata.org/wiki/Q294647}{Q294647} with two panels:
a table and a bar chart.

To demonstrate that it is possible to use other template sites and other
endpoints, I set up a template page at
\url{https://www.wikidata.org/wiki/User:Fnielsen:Synia:index} copying
a query from Wiki\-FCD and reconfigured a cloned version of Synia to use
``https://www.wikidata.org/wiki/User:Fnielsen:Synia:'' as the template
base URL and \url{https://wikifcd.wikibase.cloud/query} as the query
service URL.

\section*{Discussion/Conclusions}

The approach for the creation of new aspects and panels with Synia is more
agile and wiki-like than Scholia's method. 
While the creation of a new panel in Scholia usually involves the creation of a
new issue in GitHub, creation of a new branch, editing SPARQL and
jinja2 code, commiting, pushing, merging the branch, testing, and deploying to
Toolforge, a new panel with Synia is created by just editing a
wikipage.
Creating a new aspect with Synia can be done by creating a new
wikipage, while for Scholia it would entail editing Python code as well as
all the other steps involved in creating a panel.
Discussions about new aspects or changes in Scholia take place on
GitHub issue pages, while for Synia, discussions could take place on
the wiki, e.g., the talk page associated with the templates.

Wikis with open editing, such as Wikidata, can be vandalized and
security is an issue.
If a malicious wiki editor adds a third-party endpoint then the
browsing behavior of a Synia user will leak to the third-party site.
The problem could be alleviated by having a set of allowed endpoints,
e.g., Wikidata and Wikibase.cloud instances.

\begin{figure}[ht]
  \begin{center}
    \centerline{\includegraphics[width=\columnwidth]{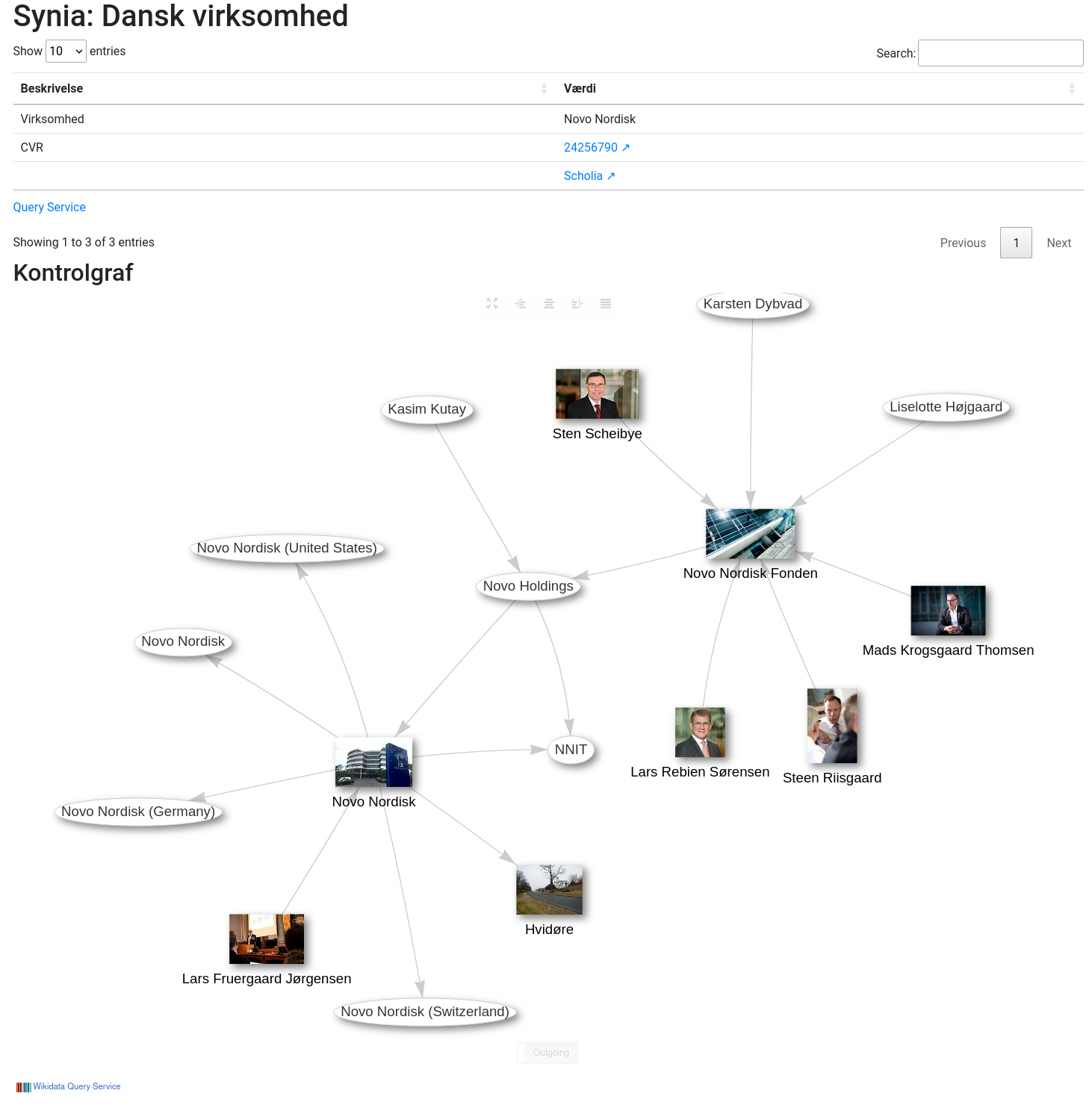}}
    \caption{Rendered page in Danish for a Danish company,
      \url{https://synia.toolforge.org/\#danskvirksomhed/Q818846},
      with a control graph panel inspired from the CVRminer
      Web application.}
    \label{fig:syniavirksomhed}
  \end{center}
\end{figure}

How language should best be handled is not clear.
Figure~\ref{fig:syniavirksomhed} shows an aspect in Danish for a
Danish company, so it is possible to control the language from
a template. However, this approach ``occupies'' a specific URI pattern and
a change of language is not possible without redoing much of the template.

Navigation with menu and search is currently missing in Synia as well as
redirects and aspect-switching that all are available in Scholia.
Instead of hardcoding such components in the Web application, it is
envisioned that components in the templates on the wiki could control
placement of menus and search forms.

SPARQL in MediaWiki templates may generate a problem as the pipe and
the equality characters in SPARQL collide with the use of the
characters to handle parameters in MediaWiki templates.
Synia's simple regular expression parsing of the wikitext does not handle
``\{\{!\}\}'' that may be used to escape the pipe character in a
MediaWiki template.
A more elaborate parsing may be needed.

\section*{Acknowledgment}

Thanks to the Scholia team, particular Daniel Mietchen
and Egon Willighagen, for continued inspiration.

% python -m scholia.tex write-bib-from-aux Nielsen2023Synia.aux
\bibliographystyle{wws2023} % Please do not change the bibliography style
% \bibliography{Nielsen2023Synia,Nielsen2023Synia_extra}

\clearpage

\begin{table}[tb]
  \begin{center}
    \begin{tabular}{|l|l|}
      \hline \bf Template & \bf URI fragment example \\ \hline
      \href{https://www.wikidata.org/wiki/Wikidata:Synia:index}{index} & \href{https://synia.toolforge.org/}{Main page} \\ 
      \href{https://www.wikidata.org/wiki/Wikidata:Synia:author-index}{venue-index} & \href{https://synia.toolforge.org/#venue}{\#venue}  \\
      \href{https://www.wikidata.org/wiki/Wikidata:Synia:author}{author} & \href{https://synia.toolforge.org/\#author/Q18618629}{\#author/Q18618629} \\
      \href{https://www.wikidata.org/wiki/Wikidata:Synia:venue-topic}{venue-topic} & \href{https://synia.toolforge.org/#venue/Q15817015/topic/Q2013}{\#venue/Q15817015/topic/Q2013} \\
       \href{https://www.wikidata.org/wiki/Wikidata:Synia:lexeme}{lexeme}
                          &
                            \href{https://synia.toolforge.org/#lexeme/L2310}{\#lexeme/L2310} \\
      \hline
    \end{tabular}
  \end{center}
  \caption{\label{table} Mapping between URI fragments and wikipages
    with SPARQL templates.}
  \label{tab:mapping}
\end{table}

\begin{table}[tb]
  \begin{center}
    \begin{tabular}{|l|l|}
      \hline \bf Template & \bf Handling \\ \hline
      = Heading 1 = & h1 HTML tag\\ 
      == Heading 2 == & h2 HTML tag \\
      === Heading 3 === & h3 HTML tag \\
      {-}{-}{-}{-} & hr HTML tag \\
      \{\{SPARQL \}\} & Submitted to endpoint \\
      \hline
    \end{tabular}
  \end{center}
  \caption{\label{table} Handling of components on the wikipage.}
  \label{tab:components}
\end{table}

\end{document}